\documentclass[conference]{IEEEtran}

\IEEEoverridecommandlockouts
\usepackage{amsmath}
\usepackage{color}
\usepackage{array}
\usepackage{stfloats}
\usepackage{amsthm} 
\usepackage{amssymb}
\usepackage{hhline}
\usepackage{float}
\usepackage{physics}
\usepackage{nccmath}
\usepackage{graphicx}
\usepackage[linesnumbered,ruled]{algorithm2e}
\usepackage{setspace}
\usepackage{booktabs}
\usepackage{subcaption}
\usepackage{mwe}
\usepackage{cite}
\usepackage{amsmath,amssymb,amsfonts}
\usepackage{graphicx}
\usepackage{textcomp}
\usepackage{xcolor}
\usepackage[utf8]{inputenc} 
\usepackage{kotex} 

\usepackage{algorithmicx}
\usepackage{xspace}
\def\BibTeX{{\rm B\kern-.05em{\sc i\kern-.025em b}\kern-.08em
    T\kern-.1667em\lower.7ex\hbox{E}\kern-.125emX}}
\begin{document}
\newcommand{\BfPara}[1]{{\noindent\bf#1.}\xspace}
\newcommand\mycaption[2]{\caption{#1\newline\small#2}}
\newcommand\mycap[3]{\caption{#1\newline\small#2\newline\small#3}}

\title{Quantum Scheduling for Millimeter-Wave Observation Satellite Constellation}

\author{
\IEEEauthorblockN{Joongheon Kim}
\IEEEauthorblockA{\textit{Korea University} \\
Seoul, Korea}
\and
\IEEEauthorblockN{Yunseok Kwak}
\IEEEauthorblockA{\textit{Korea University} \\
Seoul, Korea}
\and
\IEEEauthorblockN{Soyi Jung}
\IEEEauthorblockA{\textit{Korea University} \\
Seoul, Korea}
\and
\IEEEauthorblockN{Jae-Hyun Kim}
\IEEEauthorblockA{\textit{Ajou University}\\
Suwon, Korea}
}

\maketitle

\begin{abstract}
In beyond 5G and 6G network scenarios, the use of satellites has been actively discussed for extending target monitoring areas, even for extreme circumstances, where the monitoring functionalities can be realized due to the usage of millimeter-wave wireless links. This paper designs an efficient scheduling algorithm which minimizes overlapping monitoring areas among observation satellite constellation. In order to achieve this objective, a quantum optimization based algorithm is used because the overlapping can be mathematically modelled via a max-weight independent set (MWIS) problem which is one of well-known NP-hard problems. 
\end{abstract}


\section{Introduction}
The use of satellite constellation is widely and actively used in next generation wireless network system design and implementation~\cite{icnc20,jsac21liu,ictc20kim,marcuccio2019smaller}.
Especially, low earth orbit (LEO) satellites are getting a lot of attentions for various 6G applications such as target area observations~\cite{icte21kim} and flexible/robust network coverage extensions~\cite{jsac21liu}. Both of them require high-capacity satellite communications.

In order to realize the high-capacity satellite communications, (mmWave) frequencies are used in order to take care of the huge traffic demands and the service continuity requirements of next-generation 6G applications~\cite{icnc20,tvt16kim,tvt2021jung,pieee202105park,iotj17kim,6955961}. 
Thus, large-scale surveillance and target area observation can be realized~\cite{arvix20observe}.
In the observation satellite systems, having server duplicated/overlapped monitoring areas among satellites is not efficient even though millimeter-wave high-capacity communications can be realized. 
Thus, scheduling algorithms in order to minimize the overlapping monitoring areas have been actively studied and proposed, e.g.,~\cite{electronics20jung}.

The modeling of the overlapping area scheduling for observation satellite constellation can be realized with maximum weight independent set (MWIS) formulation~\cite{arvix20observe,tit09mit,16kim}, which is one of the well-known NP-hard problems~\cite{tit09mit}.
In order to approximately solve the NP-hard problems, many algorithms have been investigated. Among them, the use of message-passing algorithms is one of the well-known solutions~\cite{16kim,tit09mit}, whereas this paper proposes a new novel algorithm that utilizes quantum optimization and approximation methodologies~\cite{icoin20ai,ictc19qaoa}. 

Based on the advances in quantum optimization methodologies, many algorithms have been investigated for approximating combinatorial problems (e.g., MWIS~\cite{app20scheduling}, max-flow/min-cut~\cite{Cui_2016} and graph cut segmentation~\cite{tse2019graph}) and deep learning training/inference problems (e.g., Quantum Convolutional Neural Network (QCNN)~\cite{ictc20qcnn}, Quantum Random Access Memory (QRAM)~\cite{icoin21qram}, and Quantum Graph Recurrent Neural Network (QGRNN)~\cite{icoin21qgrnn}). 
In this paper, we design a quantum-based approximation algorithm for MWIS-based overlapping monitoring area scheduling in  observation satellite constellation.

The rest of this paper is organized as follows.
Sec.~\ref{sec:mwis} presents the formulation of MWIS scheduling in satellite observation modeling.
Sec.~\ref{sec:3} and Sec.~\ref{sec:4} describe the preliminaries of QAOA and QAOA-based MWIS scheduling for observation satellite constellation.
Sec.~\ref{sec:5} concludes this paper and presents future research directions.

\section{Maximum Weight Independent Set (MWIS) Formulation for Satellite Observation Scheduling}\label{sec:mwis}

We consider a network which consists of a  set of observation areas~\cite{16kim}.
According to the high data transmission rate in millimeter-wave wireless links among satellites, the transmission queue backlog in satellites can be filled in an instant, with the observation image data via synthetic aperture radar (SAR)~\cite{icte21kim}.
For the scheduling of observation satellite constellation, a conflict graph is constructed with the set of nodes (physically, observation areas) and edges where two nodes are connected by an edge if the corresponding observation areas are overlapped more than threshold among adjacent observation satellites. 
The edges between node $s_{i}$ (observation area in satellite $i$) and node $s_{j}$ (observation area in satellite $j$) of the conflict graph, i.e., $E_{(i,j)}$, can be modelled as,

\begin{equation}
E_{(i,j)} = 
\left\{
    \begin{array}{ll}
        1, & \text{if $s_{i}$ is overlapped with $s_{j}$ where }\\
              & \text{$s_{i}\in S$, $s_{j}\in S$, and $i\neq j$},\\
             0, & \text{otherwise},
        \end{array}
    \right.
    \label{eq:adj-e}
\end{equation}
where $S$ stands for the set of nodes (i.e., observation areas of satellites).

For scheduling problems, the main objective is to find the set of nodes (i.e., observation areas in satellite constellation) where two adjacent nodes those are connected via an edge cannot be simultaneously selected because it is not allowed to have huge overlapping monitoring areas among observation satellites.
This situation is obviously equivalent to the case which maximizes the summation of weights of all independent sets in a given conflict graph. Note that the weight is defined as the degree of overlapping or the number of observation data in satellite constellation. 
Thus, it is obvious that this scheduling problem can be modelled with the form of MWIS as~\cite{16kim},

\begin{eqnarray}
    \max: & & \sum_{\forall s_{k}\in S} w_{k} \cdot \mathcal{I}_{k}, \label{obj_f} \\
    \text{s.t.} & & \mathcal{I}_{i}+\mathcal{I}_{j}+E_{(i,j)}\leq 2, \forall s_{i}\in S, \forall s_{j}\in S, \label{constraint_begin} \\
    & & \mathcal{I}_{i}\in\{0,1\}, \forall s_{i}\in S,
\end{eqnarray}    
where $w_{k}$ stands for the weight of satellite $k$ (a positive integer), and
\begin{equation}
\mathcal{I}_{i} = \left\{
    \begin{array}{ll}
        1, & \text{if $s_{i}$ is scheduled where $s_{i}\in S$}, \\
        0, & \text{otherwise},
        \end{array}
    \right. \label{constraint_end}
\end{equation}
where this formulation aims at the case where conflicting links are not scheduled simultaneously. $\mathcal{I}_{i}+\mathcal{I}_{j}\leq 2$ when $E_{(i,j)}=0$ (no edge between $s_{i}$ and $s_{j}$), i.e., both of $\mathcal{I}_{i}$ and $\mathcal{I}_{j}$ can be $1$. 
On the other hand, $\mathcal{I}_{i}+\mathcal{I}_{j}\leq 1$ when $E_{(i,j)}=1$, i.e., both of $\mathcal{I}_{i}$ and $\mathcal{I}_{j}$ can not be $1$. Thus, one of them will be selected or both of them will not be selected.

\section{Preliminaries of Quantum Optimization}\label{sec:3}

This section presents the preliminaries of quantum optimization, i.e.,bra-ket notation (refer to Sec.~\ref{sec:3-1}), quantum gates (refer to Sec.~\ref{sec:3-2}), and quantum approximate optimization algorithm (QAOA) (refer to Sec.~\ref{sec:3-3}). 

\subsection{Bra–Ket Notation}\label{sec:3-1}
In quantum computing research, a bra–ket notation is widely and generally used for mathematically presenting quantum states or qubit states~\cite{app20scheduling}.
The \textit{ket} and \textit{bra} in this bra-ket notation can be represented as column vectors and row vectors.
As a result, single qubit states (i.e., $\ket{0}$ and $\ket{1}$), can be mathematically presented as,

\begin{eqnarray}
    \ket{0} &=&    
    \begin{bmatrix}
    1\\
    0\\
    \end{bmatrix},\\
    \ket{1} &=&    
    \begin{bmatrix}
    0\\
    1\\
    \end{bmatrix},
\end{eqnarray} 
and therefore,
\begin{eqnarray}
 \ket{0}&=&{\bra{0}}^{\dagger} =
    \begin{bmatrix}
    1&0
    \end{bmatrix}^{\dagger}, \\
\ket{1}&=&{\bra{1}}^{\dagger}
    \begin{bmatrix}
    0&1
    \end{bmatrix}^{\dagger}. 
\end{eqnarray} 
where $\dagger$ stands for Hermitian transpose.
Therefore, the superposition state of a single qubit state can be presented as,
\begin{equation}
    c_1\ket{0}+c_2\ket{1}=
    \begin{bmatrix}
    c_1\\
    c_2\\
    \end{bmatrix},
\end{equation}
where $c_1$ and $c_2$ are probability amplitudes, and note that the $c_1$ and $c_2$ are complex numbers~\cite{duarte}.

\subsection{Quantum Gates}\label{sec:3-2}
This section presents quantum gates or operators which mathematically represent single-qubit or $2$-qubit operations~\cite{duarte}.
First, Hadamard gate $H$, Pauli-$X$ gate $X$, Pauli-$Y$ gate $Y$, and Pauli-$Z$ gate $Z$ can be formulated as,
\begin{eqnarray}
    H &=& \frac{1}{\sqrt{2}}   
    \begin{bmatrix}
    1&1\\
    1&-1\\
    \end{bmatrix},\label{eq:Hadamard}
    \text{  } \\
    X &=&    
    \begin{bmatrix}
    0&1\\
    1&0\\
    \end{bmatrix},
    \text{  } \\
    Y &=&    
    \begin{bmatrix}
    0&-i\\
    i&0\\
    \end{bmatrix}, 
    \text{ and } \\
    Z &=&    
    \begin{bmatrix}
    1&0\\
    0&-1\\
    \end{bmatrix}. 
\end{eqnarray}

Based on this, the rotation-$X$ gate $R_{X}(\theta)$, the rotation-$Y$ gate $R_{Y}(\theta)$, and the rotation-$Z$ gate $R_{Z}(\theta)$ are as,
\begin{eqnarray}
    R_{X}(\theta) &=& 
    \begin{bmatrix}
    \cos{\frac{\theta}{2}}&-i\sin{\frac{\theta}{2}}\\
    -i\sin{\frac{\theta}{2}}&\cos{\frac{\theta}{2}}\\
    \end{bmatrix}, \\
    R_{Y}(\theta) &=& 
    \begin{bmatrix}
    \cos{\frac{\theta}{2}}&-\sin{\frac{\theta}{2}}\\
    \sin{\frac{\theta}{2}}&\cos{\frac{\theta}{2}}\\
    \end{bmatrix},
    \text{ and } \\ 
    R_{Z}(\theta) &=& 
    \begin{bmatrix}
    e^{-i\frac{\theta}{2}}&0\\
    0&e^{i\frac{\theta}{2}}\\
    \end{bmatrix},
\end{eqnarray}
where $\theta$ is an angular value.
%

\begin{figure*}[t!]
\begin{eqnarray}
    f(y) &\triangleq&f(y_1,y_2,...,y_n), \label{eq:org_fy}\\
    H_P\ket{y} &\triangleq&f(y)\ket{y}, \label{eq:org_HP}\\
    H_M &\triangleq&\sum_{k=1}^{n} X_k, \label{eq:org_HM}\\
    \ket{\gamma, \beta} &\triangleq& {e^{-i\beta_p H_M} e^{-i\gamma_p H_P} \cdots e^{-i\beta_2 H_M} e^{-i\gamma_2 H_P}e^{-i\beta_1 H_M} e^{-i\gamma_1 H_P} \ket{s},} \label{eq:state}
\end{eqnarray}
\hrulefill
\end{figure*}

\subsection{Quantum Approximate Optimization Algorithm (QAOA)}\label{sec:3-3}
QAOA is one of the widely known noisy intermediate-scale quantum (NISQ) optimization algorithms for approximating combinatorial problems~\cite{ictc19qaoa,Preskill_2018,PhysRevX.10.021067}.
This QAOA is used for formulating $H_P$ (i.e., problem Hamiltonian) and $H_M$ (i.e., mixing Hamiltonian) based on the objective function $f(y)$. Then, the QAOS generates the parameterized states $\ket{\gamma, \beta}$ by alternately and iteratively applying the $H_P$ and $H_M$ on initial state $\ket{s}$.
Here, $f(y)$, $H_P\ket{y}$, $H_M$, and $\ket{\gamma, \beta}$ are defined as~\eqref{eq:org_fy}, \eqref{eq:org_HP}, \eqref{eq:org_HM}, and \eqref{eq:state}, where $n\in\mathbb{Z}^{+}$, $p\in\mathbb{Z}^{+}$, and $X_{k}$ is the Pauli-$X$ operator applying on the $k$-th~qubit; $\gamma$ and $\beta$ are the hyper-parameters those can be computed via approximation.
Here, $H_P$ encodes $f(y)$ in~\eqref{eq:org_HP}, operating diagonally in $n$-qubit quantum basis states~\cite{Hadfield_2019}.
In the computation procedure of QAOA, via the iterative measurement of $\ket{\gamma, \beta}$, the expectation of $H_P$ should be obtained. Finally, the samples of $f(y)$ can be obtained as~\cite{ictc19qaoa},

\begin{equation}\label{exp_f}
    \expval{f(y)}_{\gamma, \beta} = \expval{H_P}{\gamma, \beta}.
\end{equation}

The near-optimal or optimal approximation values of the hyper-parameters $\gamma$ and $\beta$ are obtained using conventional optimization, e.g., stochastic gradient descent~\cite{NIPS2010_abea47ba, 4021427}.
Thus, the solution can be computed from~\eqref{exp_f} via the obtained hyper-parameters $\gamma$ and $\beta$. 
Finally, it can be shown that the QAOA-based approximation is one of widely known hybrid quantum-classical optimization algorithms where the efficient Hamiltonian design (for quantum approach) and the approximation of efficient hyper-parameters (for conventional optimization approach) are correlated~\cite{streif2019training,PhysRevA.97.022304}.

\section{Quantum Scheduling for MWIS-based Formation in Satellite Constellation}\label{sec:4}

This section consists of the design of Hamiltonian, i.e., Problem Hamiltonian, i.e., $H_P$ (refer to Sec.~\ref{sec:4-1}) and Mixing Hamiltonian, i.e., $H_M$ (refer to Sec.~\ref{sec:4-2}). Lastly, QAOA iterative computation procedure is described in Sec.~\ref{sec:4-3}.

\subsection{Problem Hamiltonian, $H_P$}\label{sec:4-1}
The problem Hamiltonian $H_P$ is designed via the linear combination of the objective Hamiltonian $H_O$ and the constraint Hamiltonian $H_C$.
The objective function and constraints in the mathematical problem for solving the considering MWIS-based scheduling problem are in $H_O$ (refer to Sec.~\ref{sec:4-1-1}) and $H_C$ (refer to Sec.~\ref{sec:4-1-2}).

\subsubsection{Hamiltonian for MWIS Objective, $H_{O}$}\label{sec:4-1-1}
Suppose that a basic Boolean function $B_1(x)=x$ exists where $x\in \{0,1\}$. According to the quantum Fourier expansion of this $B_1(x)=x$, it can be mapped to Boolean Hamiltonian $H_{B_1}$ where $I$ and $Z$ are identity operator and Pauli-$Z$ operator~\cite{hadfield2018representation}, i.e.,
\begin{equation}\label{eq:boolean_H}
    H_{B_1}=\frac{1}{2}(I-Z),
\end{equation}
therefore, the objective function~\eqref{obj_f} can be mapped into the following Hamiltonian,
%
\begin{equation}\label{H_O'}
    H_{O'}=\sum_{\forall s_{k}\in S}  \frac{1}{2} w_k (I- Z_k),
\end{equation}
where $Z_{k}$ is the Pauli-$Z$ operator applied to $\mathcal{I}_{k}$. Because the objective function~\eqref{obj_f} is mapped to $H_{O'}$, it should be maximized via the main objective of MWIS. Thus, it is obvious that this $H_{O'}$ should be maximized as well.
Therefore, the objective Hamiltonian $H_O$ should be minimized is as,
\begin{equation}\label{eq:H_O}
\boxed{
    H_{O}=\sum_{\forall s_{k}\in S}  \frac{1}{2}w_k Z_k}.    
\end{equation}

\subsubsection{Hamiltonian for MWIS Constraints}\label{sec:4-1-2}
In the MWIS-based scheduling problem, we should avoid the case where both adjacent nodes of the conflict graph are selected.
The scheduled and unscheduled nodes have states are denoted as $\ket{1}$ and $\ket{0}$. Here, $N_i$ and $N_j$ are defined as the arbitrary nodes, and $E_A(N_i, N_j)$, $E_B(N_i, N_j)$, and $E_C(N_i, N_j)$ stands for the edge notations for three cases where,
\begin{itemize}
    \item $E_A(N_i, N_j)$ for \textit{Case A}: $s_{i}$ and $s_{j}$ are not scheduled,
    \item $E_B(N_i, N_j)$ for \textit{Case B}: One of $s_{i}$ and $s_{j}$ is scheduled,
    \item $E_C(N_i, N_j)$ for \textit{Case C}: Both of $s_{i}$ and $s_{j}$ are scheduled (impossible situation).
\end{itemize}

Suppose that the weights of $N_i$ and $N_j$ in \textit{Case C} are defined as $W_{N_i}$ and $W_{N_j}$. Under this definition, the constraint function $C'(i,j)$, which counts the impossible situations, can be represented as,

\begin{equation}
    C'(i,j)=\sum_{i=1}^{n} \sum_{j=1}^{n}  (W_{N_i}+W_{N_j})\abs{E_C(N_i,N_j)}
\end{equation}
where $i>j$; and $n$ and $\abs{E_C(N_i,N_j)}$ are the number of nodes and the number of $E_C(N_i,N_j)$ where $i>j$. This is a primary condition for avoiding impossible situations.

Based on the mathematical program of MWIS problem formulation (i.e., \eqref{eq:adj-e}--\eqref{constraint_end}), $C'(i,j)$ can be re-formed as $C(i,j)$ as,
\begin{eqnarray}\label{eq:C}
    C(i,j)&=&\sum_{{\forall s_{i}\in S}} \sum_{{\forall s_{j}\in S}} (w_i+w_j)E_{(i,j)} \nonumber \\
    &=&\sum_{{\forall s_{i}\in S}} \sum_{{\forall s_{j}\in S}} (w_i+w_j)(\mathcal{I}_{i}\land \mathcal{I}_{j}),
\end{eqnarray}
where $i>j$; and $\land$ stands for an \textsf{AND} gate. 
In~\eqref{eq:C}, $C(i,j)$ should be $0$ because it stands for the number of impossible situations. If making this $C(i,j)$ be not possible, this $C(i,j)$ should be minimized as much as possible.
According to the quantum Fourier expansion of \textsf{AND} gate $B_2(x_1,x_2)$, it can be mapped to the following Boolean Hamiltonian $H_{B_2}$ where the $Z_{1}$ and $Z_{2}$ in this equation are the Pauli-$Z$ operators applying on $x_1$ and $x_2$, respectively~\cite{hadfield2018representation},
\begin{eqnarray}
    B_2(x_1,x_2)&=&x_1\land x_2 \text{ where }\nonumber \\
    & & x_1\in \{0,1\}\text{ and } x_2\in \{0,1\}, \label{eq:boolean_f2} \\
    H_{B_2}&=&\frac{1}{4}(I-Z_1-Z_2+Z_1 Z_2). \label{eq:boolean_H2}
\end{eqnarray}

Based on this result, the constraints~\eqref{eq:C} can be mapped into following Hamiltonian,
\begin{equation}
    H_{C'}=  \sum_{{\forall s_{i}\in S}} \sum_{{\forall s_{j}\in S}}  \frac{1}{4}(w_i+w_j)(I-Z_i-Z_j+Z_i Z_j), 
\end{equation}
where $i>j$; and $Z_{i}$ and $Z_{j}$ are the Pauli-$Z$ operators applied to $\mathcal{I}_i$ and $\mathcal{I}_j$, respectively.
Because $C(i,j)$ should be zero (or it should be minimized, as explained before), the $H_{C'}$ which is mapped from $C(i,j)$ should be minimized, as well. Thus, the constraint Hamiltonian $H_C$ is as,
\begin{equation}\label{eq:H_C}
\boxed{
    H_{C}=\sum_{{\forall s_{i}\in S}} \sum_{{\forall s_{j}\in S}}-\frac{1}{4}(w_i+w_j)(Z_i+Z_j-Z_i Z_j)},
\end{equation}
where $i>j$.

Based on the obtained $H_O$ and $H_C$ in \eqref{eq:H_O} and \eqref{eq:H_C}, the problem Hamiltonian $H_P$ is defined as,
\begin{equation}\label{eq:H_P}
    \boxed{H_P=H_O+\rho H_C}, 
\end{equation}
where $\rho$ is a hyper-parameter that represents the penalty rate which means the ratio at which $H_C$ (constraints) affects $H_P$ compared to $H_O$ (objective) ($\rho\geq 1$). 

\subsection{Mixing Hamiltonian, $H_M$}\label{sec:4-2}
The mixing Hamiltonian $H_{M}$ produces various cases which can appear in the given MWIS-formulated combinatorial problem.
Our considering MWIS-based observation scheduling problem can be formulated by a binary bit string which presents a set of nodes. Therefore, various cases can be created by flipping the state of each node, mathematically modelled as $\ket{0}$ or $\ket{1}$.
The bit-flip can be handled by the Pauli-$X$ operator. Therefore, $H_{M}$ can be formed as,
\begin{equation}\label{eq:H_M}
    \boxed{H_M=\sum_{{\forall s_{k}\in S}} X_k}.
\end{equation}

\subsection{QAOA Iterative Computation}\label{sec:4-3}
The application of the designed Hamiltonian to the QAOA iterative optimization computation sequence starts when the design of Hamiltonian functions, i.e., $H_P$ and $H_M$ in \eqref{eq:H_P} and \eqref{eq:H_M}, are completed. Then the iterative optimization computation procedure is as follows.
\begin{itemize}
    \item First of all, the parameterized state $\ket{\gamma, \beta}$ can be generated by applying $H_P$ and $H_M$ to~\eqref{eq:state}, as defined in~\eqref{eq:H_O},~\eqref{eq:H_C},~\eqref{eq:H_P}, and~\eqref{eq:H_M}. Note that the initial state $\ket{s}$ is set to the equivalent superposition state using the Hadamard gates in~\eqref{eq:Hadamard}.
    \item The expectation of $H_P$ can be measured on the generated parameterized state $\ket{\gamma, \beta}$. Here, The parameters $\gamma$ and $\beta$ are iteratively updated with traditional optimization computation procedure.
    \item When the QAOA iterative computation sequence terminates, the optimal (or approximated) parameters $\gamma_{\textsf{OPT}}$ and $\beta_{\textsf{OPT}}$ are finally obtained.
\end{itemize}

Therefore, the MWIS-based monitoring area scheduling solution in observation satellite constellation can be obtained by the measurement of the expectation of $H_P$ on the optimal state $\ket{\gamma_{\textsf{OPT}}, \beta_{\textsf{OPT}}}$ as,
\begin{equation}
    \boxed{\expval{F}=\expval{H_P}{\gamma_{\textsf{OPT}}, \beta_{\textsf{OPT}}}},
\end{equation}
where $\expval{F}$ is the expectation of the MWIS objective function~\eqref{obj_f} for the obtained solution samples.

\section{Concluding Remarks and Future Work}\label{sec:5}
In beyond 5G and 6G communication networks, satellites has been actively used in many applications such as seamless monitoring target areas, even for extreme circumstances, thanks to the use of high-capacity millimeter-wave wireless links in satellites. This paper proposes a scheduling algorithm which aims at the minimization of overlapping monitoring areas among observation satellite constellation. To achieve this goal, a quantum optimization based algorithm is used because the our considering overlapping formulation can be mathematically modelled via a max-weight independent set (MWIS) problem. 

As a future research direction, the proposed algorithm can be evaluated with various realistic satellite scenarios and TensorFlow-Quantum based software implementation~\cite{broughton2020tensorflow}.

\section*{Acknowledgment}
This work was supported by the National Research Foundation of Korea (NRF) grant funded by the Korea government (MSIT) (2019M3E4A1080391, 2021R1A4A1030775). J. Kim, S. Jung, and J.-H. Kim are the corresponding authors of this paper (e-mails: joongheon@korea.ac.kr, jungsoyi@korea.ac.kr, jkim@ajou.ac.kr).

\bibliographystyle{IEEEtran}
\bibliography{ref_sagin,ref_opt,ref_quantum,ref_aimlab}

\end{document}